\documentstyle[12pt]{article}
\title{\huge Edge wetting of an Ising three-dimensional system} 
\author{\bf  
 L. Bahmad, A. Benyoussef and H. Ez-Zahraouy$^{*}$. 
\\
 Laboratoire de Magn\'{e}tisme et de la Physique
 des Hautes Energies
\\
Universit\'{e} Mohammed V, Facult\'{e} des Sciences, Avenue Ibn Batouta,\\  
Rabat  B.P. 1014, Morocco
}
\date{ }
\begin{document}
\maketitle 

\begin{abstract}
\mbox{~~~  } The effect of edge on wetting and layering transitions of a three-dimensional spin-$1/2$ Ising model is investigated, in the presence of longitudinal and surface magnetic fields, using mean field (MF) theory and Monte Carlo (MC) simulations. 
For $T=0$, the ground state phase diagram shows that there exist only three allowed transitions, namely: surface and bulk transition, surface transition and bulk transition. 
However, there exist a surface intra-layering temperature $T_{L}^{s}$, above which the surface and the intra-layering surface transitions occur. While the bulk layering and intra-layering transitions appear above an other finite temperature $T_{L}^{b} (\ge T_{L}^{s})$. These surface and bulk intra-layering transitions are not seen in the perfect surfaces case.
 Numerical values of $T_{L}^{s}$ and $T_{L}^{b}$, computed by Monte Carlo method are found to be smaller than those obtained using mean field theory. However, the results predicted by the two methods become similar, and are exactly those given by the ground state phase diagram, for very low temperatures. 
On the other hand, the behavior of the local magnetizations as a function of the external magnetic field, shows that the transitions are of the first order type. \\
$T_{L}^{s}$ and $T_{L}^{b}$ decrease when increasing the system size and/or the surface magnetic field. In particular, $T_{L}^{b}$ reaches the wetting temperature $T_{w}$ for sufficiently large system sizes.
\end{abstract}
 
\noindent ----------------------------------- \newline
($^{*}$) Corresponding author: ezahamid@fsr.ac.ma
\newpage

\section{Introduction}
\mbox{  } Recently, much attention has been directed to study the layering transitions of magnetic solid films. Several models have been proposed to study such physical systems. A variety of possible phase transitions has been reviewed by  de Oliveira and Griffiths [1], Pandit and Wortis [2], Pandit {\it et al.} [3] and Ebner {\it et al.} [4]. Such transitions have been observed in a variety of systems including for example $^{4}He $ [5] and ethylene adsorbed on graphite [6].
Hanke {\it et al.} [7] show that symmetry breaking fields give rise to nontrivial and long-ranged order parameter profiles for critical systems such as fluids, alloys, or magnets confined to wedges.
Benyoussef and Ez-Zahraouy [8] have studied the layering transitions of a spin-$1/2$ Ising film using transfer matrix and the real space renormalization group method [9].
By using Monte Carlo simulations on thin Ising films with competing walls, Binder {\it et al.} [10], found that occurring phase transitions belong to the universality class of the two-dimensional Ising model and found that the transition is shifted to a temperature just below the wetting transition of a semi-infinite fluid. \\
On the other hand, a variation of phase diagrams with the strength of the substrate potential in
a lattice gas model for multilayer adsorption is studied by Patrykiejew 
{\it et al.} [11] using Monte Carlo simulations and molecular field
approximation. Therefore, the effect of finite size on such transitions has been studied, in thin film
confined between parallel plates or walls, by Nakanishi and Fisher [12] using
mean field theory and by Bruno {\it et al.} [13] taking into account the
capillary condensation effect. In the framework of the mean field theory, we found in a previous work [14], the wetting and layering transitions of three-dimensional spin$-1/2$ Ising transverse model in the presence of both an external and surface magnetic fields. We showed, in our previous work [15], the existence of the layering transitions for a film, with finite thickness and infinite surfaces. We have found that the wetting temperature $T_w$ depends weakly on the surface corrugation degree for fixed values of the  surface magnetic field. \\
In addition, by applying a suitable effective interface model at liquid-vapor coexistence, Rejmer {\it et al.} [16] found a filling transition at which the height of the meniscus becomes macroscopically large while the planar walls of the wedge far away from its remain non-wet up to the wetting transition occurring at $T_w$. They also showed that the discontinuous filling transition is accompanied by a pre-filling line extending into the vapor phase of the bulk phase diagram and describing a transition from a small to a large, but finite, meniscus height. 
Long-ranged order parameter profiles, for critical systems such as fluids, alloys, or magnets confined to wedges, have been studied by Hanke {\it et al.} [7]. They also discuss the properties of the corresponding universal scaling functions of the order parameter profile and the two-point correlation function, and determine the critical exponents for the normal transition. \\
\mbox{  } The aim of this work is to study the effect of edge surfaces on the wetting and layering transitions of a three-dimensional spin$-1/2$ Ising model in the presence of external and surface magnetic fields, using  mean field theory (MF) and Monte Carlo (MC) simulations. It is found that there exist a surface intra-layering temperature $T_{L}^{s}$, above which the surface and the intra-layering surface transitions occur. While the bulk layering and intra-layering transitions appear above an other finite temperature $T_{L}^{b} (\ge T_{L}^{s})$. Such results were not seen neither in the perfect surfaces case [2,3,8,9,11,14,15] nor in the continuous model case [7,16].
The outline of this paper is as follows. In the following section we describe the model and the formulations used. Section 3 is devoted to results and discussions. \\

\section{Model and method }
The system we are studying, Fig. 1, is formed with $N$ layers. Each layer '$k$' ($k=1,2,...,N$), is formed with two perpendicular perfect planes and 
contains $2(N-k)+1$ spin chains which are infinite in the $y-$direction.
A uniform surface magnetic field $H_{s}$ is applied on the planes $(x,y,z=1)$ , $(x=1,y,z)$ of the layer $k=1$. However, the system is invariant by translation, in the $y-$direction, so the coordinate $y$ will be dropped in the following.
An external field $H$ is applied on the global system.  
The Hamiltonian governing the system, can be written as 
\begin{equation}
{\cal H}=-\sum_{<i,j>}J_{ij}S_{i}S_{j} -\sum_{i}H_{i}S_{i}
\end{equation}
where, $S_{i}=\pm 1$ are spin$-1/2$ Ising random variables, and $J_{ij}=J$ are the exchange interactions assumed to be constant. The system is assumed to be infinite in the direction $y$, so the variable $y$ will be cancelled in all the following. $H_{i}$ is the total
magnetic field applied on a spin located on a site 'i' of coordinates $(x,z)$. 
 defined by: 
\begin{equation}
H_{i}=H(x,z)=\left\{ 
\begin{array}{lll}
H+H_{s} & \mbox{for} & x=1, z=1,...,N \\ 
H+H_{s} & \mbox{for} & z=1, x=1,...,N \\
H & \mbox{~~~} &  elsewhere  \\ 
\end{array}
\right.
\end{equation}
$H$ and $H_{s}$ are, respectively, the external and surface magnetic fields
both applied in the $z-$direction. \\
Using the mean field theory, we compute the magnetizations and the free energy of such a system. However, the magnetizations can be expressed as:
\begin{equation}
m(x,z)=\tanh(\beta(2m(x,z)+m(x,z+1)+m(x,z-1)+m(x+1,z)+m(x-1,z)+H(x,z)))
\end{equation}
where $\beta=1/(k_{B}T)$, $k_{B}$ denotes the Boltzmann constant and $T$ the absolute temperature. \\ 
Eq.(3) of the magnetizations is the Callen identity with an external field $H(x,z)$ so that  
the total free energy of the system can be derived as :
\begin{equation}
\begin{array}{l}
F[m(x,z)]=\sum_{x,z}\{T[\frac{1-m(x,z)}{2}Log(1-m(x,z)) 
+\frac{1+m(x,z)}{2}Log(1+m(x,z))]   \\
-\frac{J}{2}m(x,z)[2m(x,z)+m(x,z+1)+m(x,z-1)+m(x+1,z)+m(x-1,z)] \\
-m(x,z)H(x,z)
\}.
\end{array}
\end{equation}
The system is with free boundary conditions. \\

\section{Results and discussion} 
The notation ($1^{p}1_{q}$) with $p=0,1,2,...,N$ and $q=0,1,3,5,...,2(N-p)-1$, will be used to denote that the first $p$ layers and the $q$ first spin chain of the layer $p+1$are in a magnetic state "up"; while the remaining $N-(p+1)$ layers are in the state "down". In particular, the notation $1^{N}$ (resp. $O^{N}$) will be used to denote a configuration with positive magnetization for all layer spins (resp. negative magnetization for all layer spins) of the system. Fig. 2 illustrates an example of these notations for $N=4$. \\
The ground state energy of a given configuration is calculated exactly from the Hamiltonian $(1)$. 
The transition from a configuration $(1^{p}1_{q})$ to an other configuration $(1^{p'}1_{q'})$ is obtained by the equality of their energies.
It is found that the ground state transitions, illustrated by Fig. 3, occur according to linear forms $H/J = a + b(H_{s}/J)$. 
We found three ground state transitions: \\
$\bullet$ The surface transition $O^{N} \leftrightarrow (1^{1}1_{0})$ located at $H/J=-H_{s}/J+2(N-1)/(2N-1)$. \\
$\bullet$ The bulk transition $(1^{1}1_{0}) \leftrightarrow 1^{N}$ occurring at $H/J=2/(1-N)$. \\
$\bullet$ The bulk and surface transition $O^{N} \leftrightarrow 1^{N}$  localised at $H/J=((1-2N)/(N^2))H_{s}/J$. It is clear that for large values of the system size $N$, ($N \rightarrow \infty$), the bulk and surface, and the bulk transitions are shifted to $H/J=0$. Whereas the surface transition is  located at $H/J=-H_{s}/J+1$. This is in good agreement with the results we have established in our previous works [13,15], for the infinite perfect surfaces case. \\
\mbox{~~~} Although the established equation are valid for an arbitrary system size $N$, numerical results are given for two system sizes $N=4$ (thin film) and $N=20$ (thick film) spins in both directions $x$ and $z$.
In order to examine the effect of temperature on wetting and layering transitions, we plot in Fig. 4 the corresponding phase diagram, for a fixed value of the surface magnetic field $H_{s}$, by using the mean field (MF) theory. Fig. 4 is plotted for $H_{s}/J=1.0$ and a small system size $N=4$. From the ground state phase diagram (see Fig. 3), the situation corresponding to Fig. 4 is located in the region: bulk transition. Indeed, for very small temperature values, the only transition seen is $O^{4} \leftrightarrow 1^{4}$. While the increasing temperature leads to the intra-layering surface transitions and the intra-layering bulk transitions as well as the inter-layering transitions. In the case of $N=4$, Fig. 2 shows an example of the intra-layering transitions, which correspond to the change of the spin chains from  "down" to "up" state inside the layer (in surface and bulk).
One can note that these transitions are due to the geometry of the surfaces, independently on the system size $N$. Indeed, the intra-layering surface transitions   are: $O^{4} \leftrightarrow (1^{0}1_{1})$, $(1^{0}1_{1}) \leftrightarrow (1^{0}1_{3})$, $(1^{0}1_{3}) \leftrightarrow (1^{0}1_{5})$, $(1^{0}1_{5}) \leftrightarrow (1^{1}1_{0})$ and $(1^{1}1_{0}) \leftrightarrow 1^{4}$; while the intra-layering  bulk  transitions are  $O^{4} \leftrightarrow (1^{1}1_{1})$, $(1^{1}1_{1}) \leftrightarrow (1^{1}1_{3})$, $(1^{1}1_{3}) \leftrightarrow (1^{2}1_{0})$, $(1^{2}1_{0}) \leftrightarrow (1^{2}1_{1})$, $(1^{2}1_{1}) \leftrightarrow (1^{3}1_{0})$ and $(1^{3}1_{0}) \leftrightarrow 1^{4}$ (see Fig. 2). \\
 The profile of the magnetizations $m(1,1)$, $m(1,2)$, $m(2,2)$, $m(2,3)$  are given in Figs. 5a and 5b.  From these figures, it is clear that the configuration $(1^{0}1_{1})$ is seen ( Fig. 5a) before the configuration $(1^{1}1_{3})$. The same argument is valid for the configurations $(1^{1}1_{1})$ and $(1^{1}1_{3})$: Fig. 5b. \\
   On the other hand, the Monte Carlo calculations are performed on a system with a total number of spins $4\times 4\times 100=1 600$ ($N=4$ and $ n_y=100$ spins in the $y$-direction).
The phase diagram obtained using the Monte Carlo (MC) method is illustrated by Fig. 6 for the same surface magnetic field value used in the mean field method $H_{s}/J=1.0$ (Fig. 4). Comparing these figures. (Fig. 4 and Fig. 6), plotted for the same surface magnetic field value, it is seen that the layering transitions are found by the (MC) method, when increasing the temperature, before these same transitions can be seen by the (MF) method. But the global transitions are found by the two methods, and qualitatively the phase diagrams obtained exhibit similar topologies. It is worth to note that for very low temperature, (MF) and the (MC) are in good agreement with the ground state phase diagram. \\
In order to outline the above results, we plot the local magnetizations $m(1,1)$, $m(1,2)$, $m(2,2)$ and $m(2,3)$ in Figs. 7a and 7b, as a function of the reduced bulk field $H/J$. 
  When increasing the external field $H/J$ the corners transits (Figs. 7a and 7b) before the other intra-layering transitions. Such behavior was not obtained in the perfect surface case model. 
 However, the other layer-by-layer transitions occur when increasing the external field at specific temperature values. 
As one can expects, the external field values needed to make arising the intra-layering and inter-layering transitions  increase with increasing the order of the layer counted  from the surface $k=1$. This is qualitatively in a good agreement with our previous works [13-15].  \\
The phase diagram of the intra-layering and layer-by-layer transitions, in the plane $(H/J,T/J)$ for a larger system size with $N=20$ and $H_{s}/J=1.0$, is plotted in Fig. 8, by using the mean field method. Comparing this figure with Fig. 4, it is seen that the increasing system size effect is to decrease the intra and inter layering temperature values. The bulk layering transitions are shifted to higher external magnetic field values, for a fixed surface magnetic field value. \\
To complete this study, we have investigated in Fig. 9 the surface, $T_{L}^{s}$, and bulk $T_{L}^{b}$ intra-layering temperature profiles as a function of the system size $N$ for two surface magnetic field values $H_{S}/J=1.0$ and $H_{S}/J=0.9$. It is found that $T_{L}^{s}$ (Fig. 9a) (as well as $T_{L}^{b}$, Fig. 9b) decreases when increasing the system size. For a fixed system size $N$, these intra-layering temperatures decrease when increasing the surface magnetic field values. Moreover, $T_{L}^{b}$ reaches the wetting temperature for sufficiently large values of the system size. \\
To conclude this study, we note that the presence of an edge (angle $\pi /2$) decreases the wetting temperature compared with the perfect case (angle $\pi$), see ref. [14]. This finding is in a good agreement the results obtained by Rejmer {\it et al.} [16].

\section{Conclusion}
The intra and inter-layering transitions of a three-dimensional spins Ising model with edge surfaces are studied within the mean field theory and Monte Carlo simulations. At $T=0$, the only layering transitions, occurring under the effect of an external and a surface magnetic fields, are: surface transition, bulk transition and surface to bulk transition.  
When increasing the temperature, a succession of intra-layering transitions, absent in the case of perfect surfaces [13,15] and the continuous model [7,16], are found.  
The layering temperatures obtained by the mean field theory are higher than those predicted by MC method. The two methods lead to similar results for very low temperatures. The effect of the edge on the behavior of the local magnetizations $m(x,z)$, as a function of the reduced bulk field $H/J$, is also investigated. We have analyzed the effect the system size on the surface and bulk layering temperatures as well as on the wetting temperature.

\section{Aknowledgement}
The authors would like to thank S. Dietrich for helpful discussions.
 
\newpage \noindent{\bf References}

\begin{enumerate}
\item[[1]] M. J. de Oliveira  and R. B. Griffiths , Surf. Sci. {\bf 71}, 687 (1978).

\item[[2]] R. Pandit and M. Wortis, Phys. Rev. {\bf B 25}, 3226 (1982).  

\item[[3]] R. Pandit, M. Schick and M. Wortis, Phys. Rev. {\bf B 26}, 8115 (1982).

\item[[4]] C. Ebner, W. F. Saam and A. K. Sen, Phys. Rev. {\bf B 32}, 1558 (1985).

\item[[5]] S. Ramesh, Q. Zhang, G. Torso and J. D. Maynard, Phys. Rev. Lett. {\bf 52},2375 (1984). 

\item[[6]] M. Sutton, S. G. J. Mochrie  and R. J. Birgeneau, Phys. Rev. Lett. {\bf 51},407 (1983);\\
S. G. J. Mochrie, M. Sutton, R. J. Birgeneau, D. E. Moncton and P. M. Horn, Phys. Rev. {\bf B 30}, 263 (1984).

\item[[7]] A. Hanke, M. Krech, F. Schlesener and S. Dietrich, Phys. Rev. E {\bf 60}, 5163 (1999).

\item[[8]] A. Benyoussef and H. Ez-Zahraouy  Physica {\bf A 206}, 196 (1994).

\item[{[9]}]  A. Benyoussef and H. Ez-Zahraouy, J. Physique  {\bf I 4}, 393 (1994).

\item[[10]] K. Binder, D. P. Landau and A. M. Ferrenberg, Phys. Rev. Lett. {\bf 74}, 298 (1995); Phys. Rev. E {\bf 51}, 2823 (1995). 

\item[[11]] A. Patrykiejew A., D. P. Landau and K. Binder, Surf. Sci. {\bf 238}, 317 (1990).

\item[[12]] H. Nakanishi and M. E. Fisher, J. Chem. Phys. {\bf 78},3279  (1983).

\item[[13]] E. Bruno, U. Marini, B. Marconi and R. Evans, Physica A {\bf 141A}, 187 (1987).

\item[{[14]}]  L. Bahmad, A. Benyoussef, A. Boubekri and H. Ez-Zahraouy, 
Phys. Stat. Sol. (b) {\bf 215}, 1091 (1999).

\item[{[15]}]  L. Bahmad, A. Benyoussef and H. Ez-Zahraouy, Physica  {\bf A 303}, 525 (2002).


\item[[16]] K. Rejmer, S. Dietrich and M. Napiorkowski, Phys. Rev. E {\bf 60}, 4027 (1999).


\end{enumerate}
\newpage \noindent{\bf Figure Captions}\newline
\newline
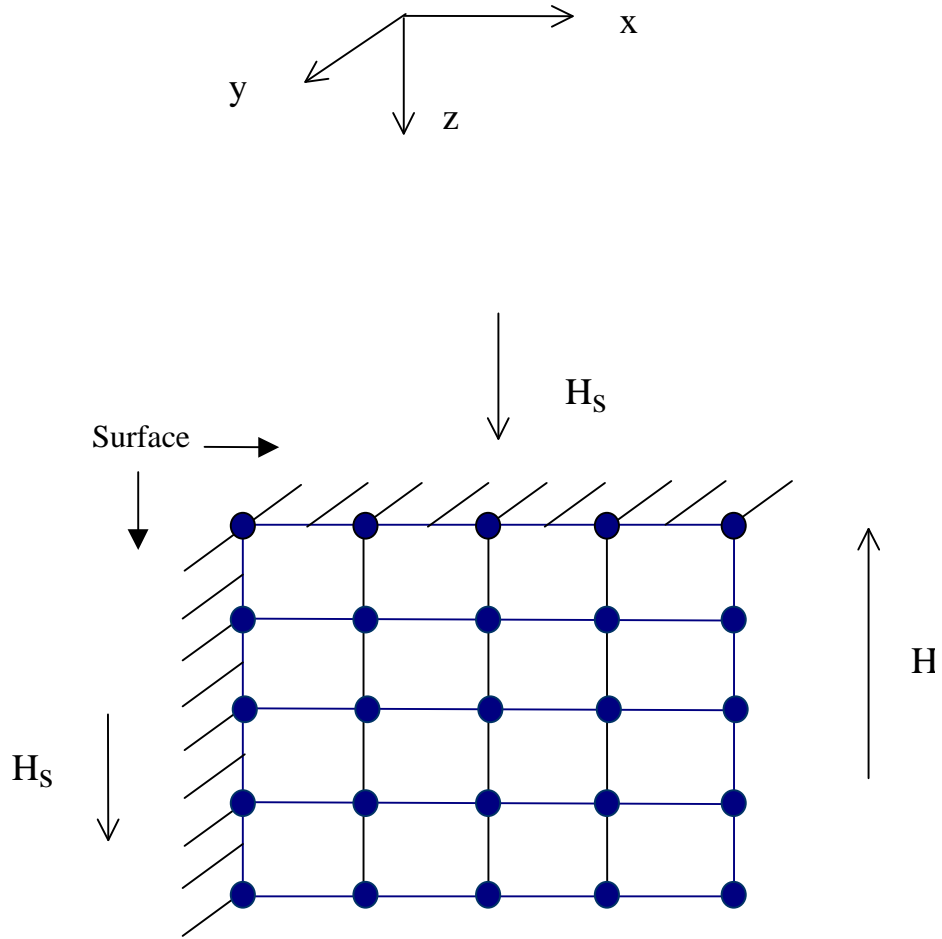
\noindent{\bf Figure 1}: Geometry of the system formed with two surfaces $(x,y,z=1)$ and $(x=1,y,z)$ with $N$ spins in both the $x$ and $z-$directions. The system is infinite in the $y-$direction. A uniform surface magnetic field $H_{s}$ is applied on the planes $(z=1,x,y)$ and $(x=1,y,z)$. An external magnetic field $H$ is applied to the global system. 

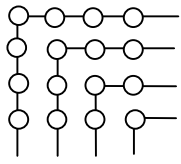
\noindent{\bf Figure 2}: Sketch of different possible configurations for a system with $N=4$. Symbols ($\circ$) and ($\bullet$)  correspond to spin "down" and spin "up" respectively. 
The notation ($1^{p}1_{q}$) where $p=0,1,2,...,N$ and $q=0,1,3,5,...,2(N-p)-1$, will be used to denote that the first $p$ layers and the $q$ first spins of the layer $p+1$  are with positive magnetizations, while the remaining $N-(p+1)$ layers are with negative magnetizations. In particular, the notation $1^{N}$ (resp. $O^{N}$) will denote a configuration in the state "up" for all layer spins (resp. state "down" for all layer spins) of the system. \\

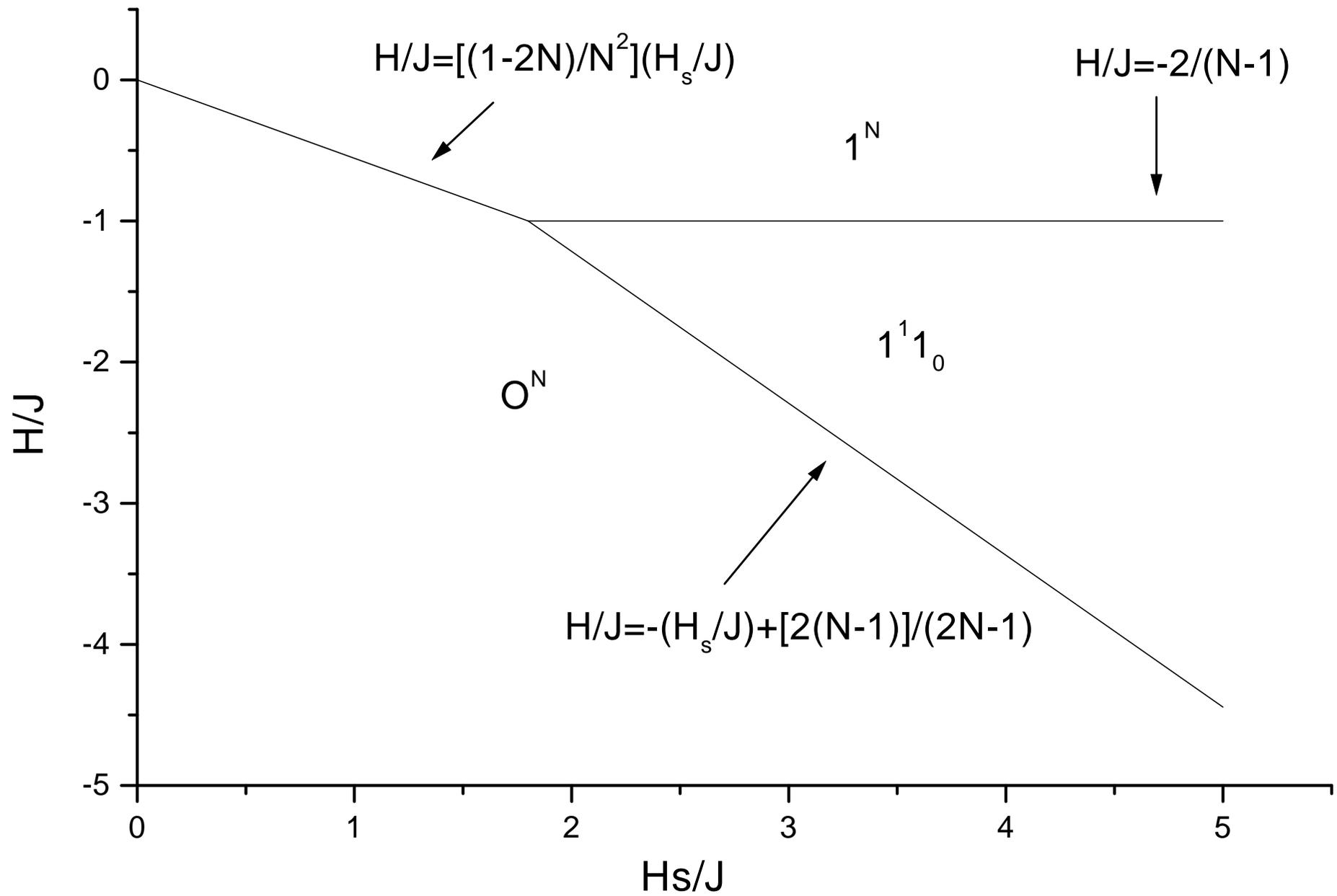
\noindent{\bf Figure 3}: The ground state phase diagram in the $(H/J,H_{s}/J)$ plane. There exist only three transitions, namely: the surface layering transition $O^{N} \leftrightarrow (1^{1}1_{0})$ , the surface to bulk layering transition $(1^{1}1_{0}) \leftrightarrow 1^{N}$ and the bulk layering transition $O^{N} \leftrightarrow 1^{N}$. \\

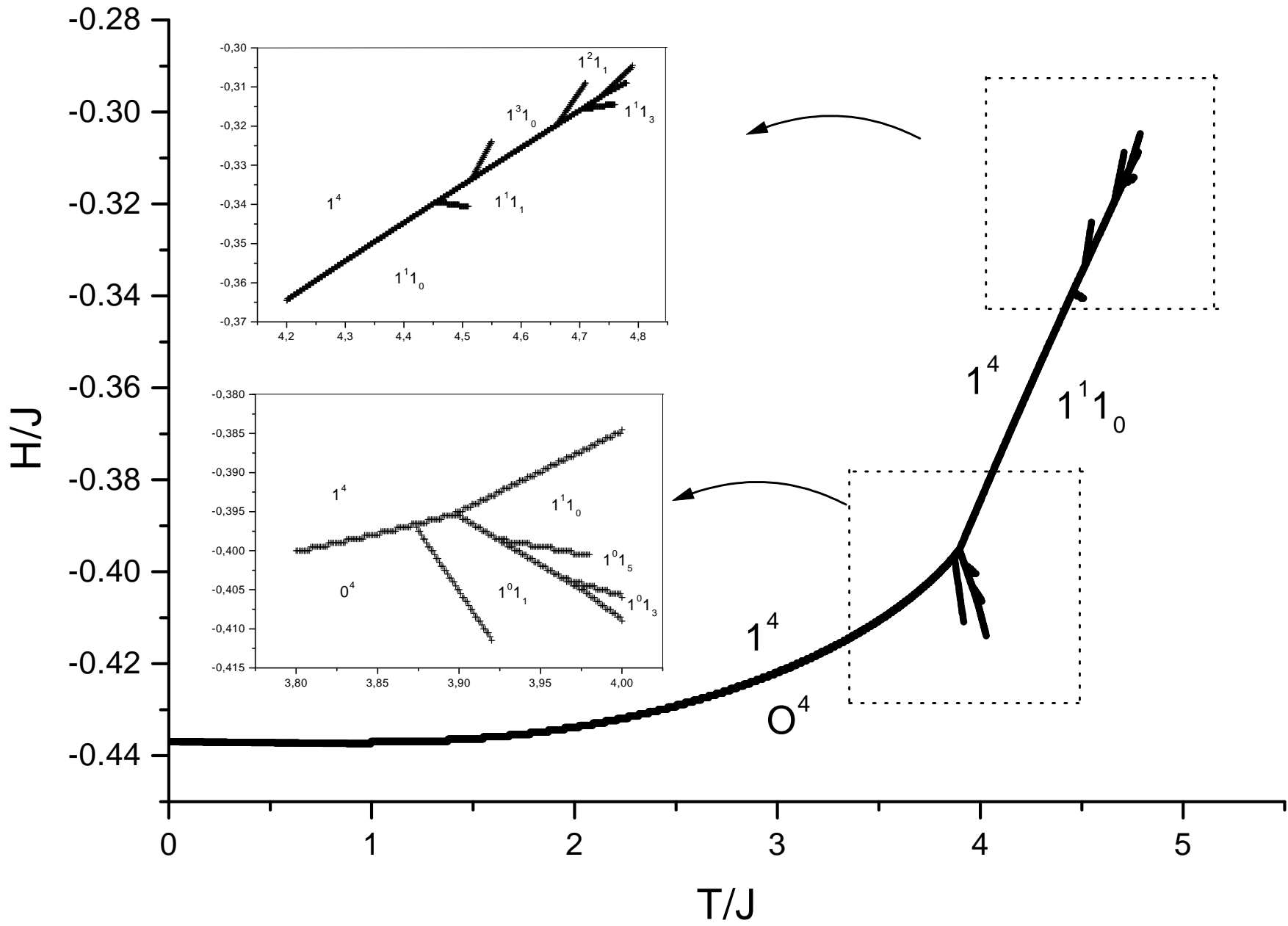
\noindent{\bf Figure 4}: Layer-by-layer and intra-layering transitions, in the plane $(H/J,T/J)$, using the mean field method for a system with $N=4$ and $H_{s}/J=1.0$. \\               

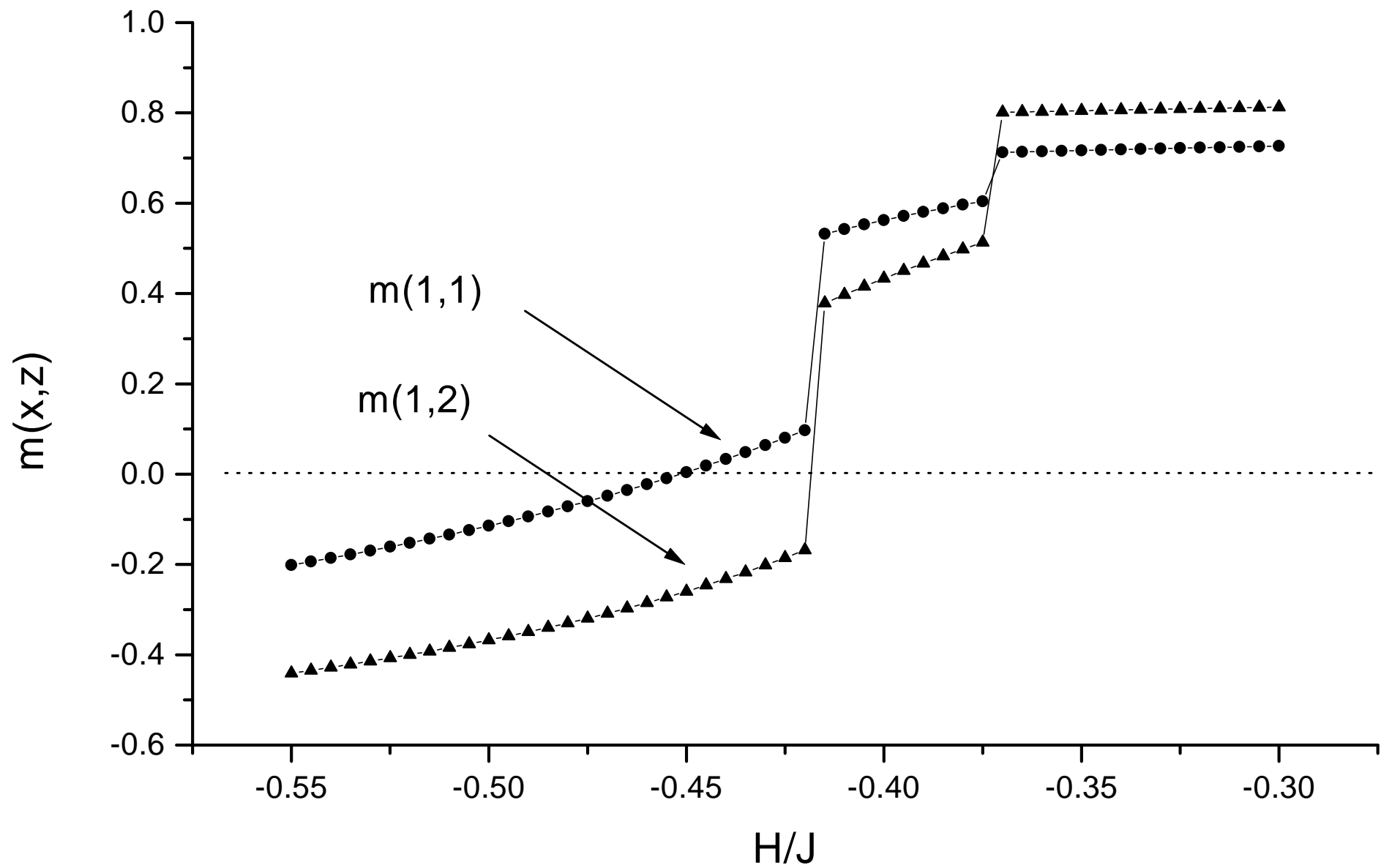
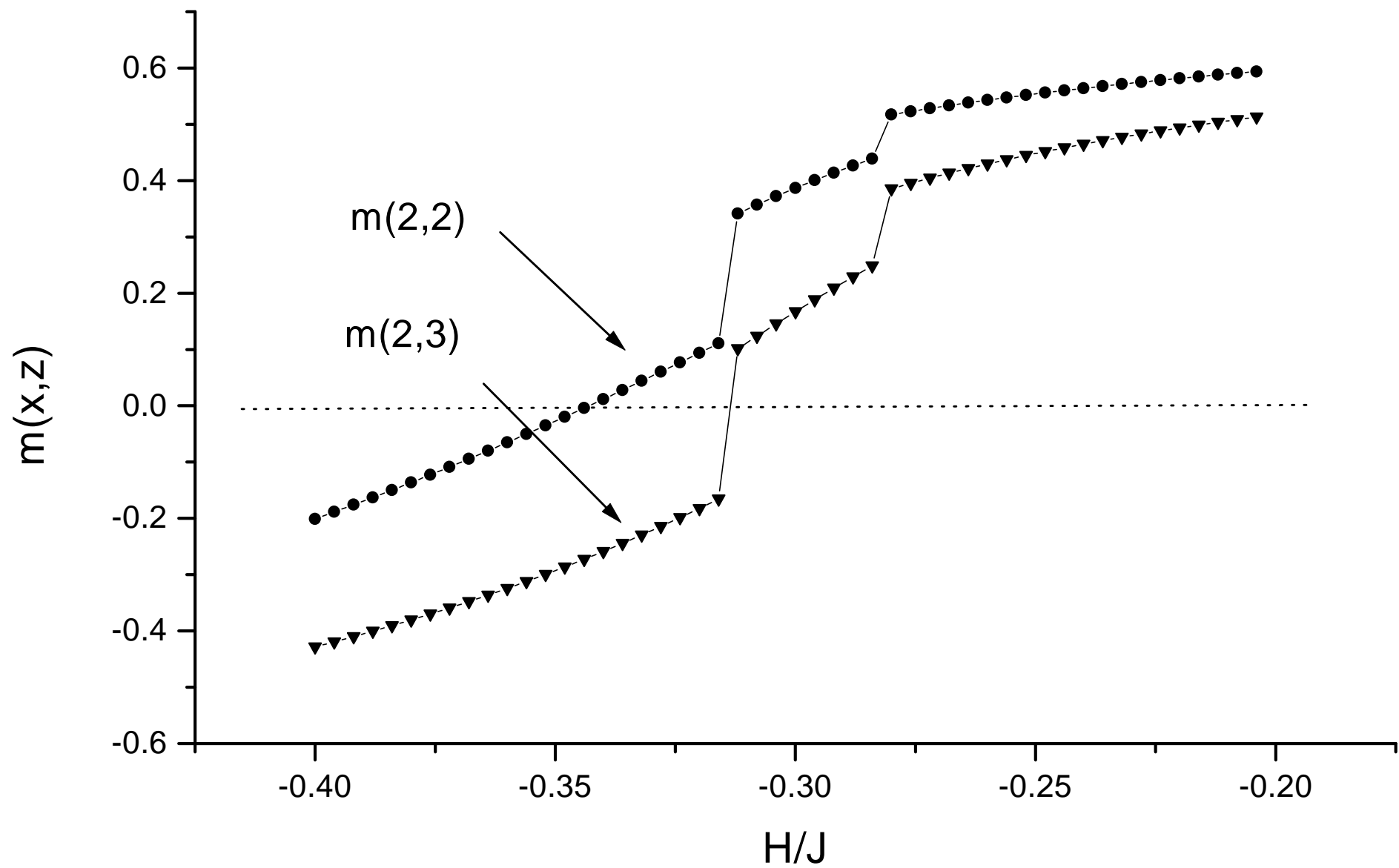
\noindent{\bf Figure 5}: Magnetization profiles as a function of the reduced bulk magnetic field $H/J$ for $N=4$ and $H_{S}/J=1.0$, of $m(1,1)$ and $m(1,2)$ at $T/J=3.95$ (a), $m(2,2)$ and $m(2,3)$ at $T/J=4.75$ (b), by the mean field method. \\

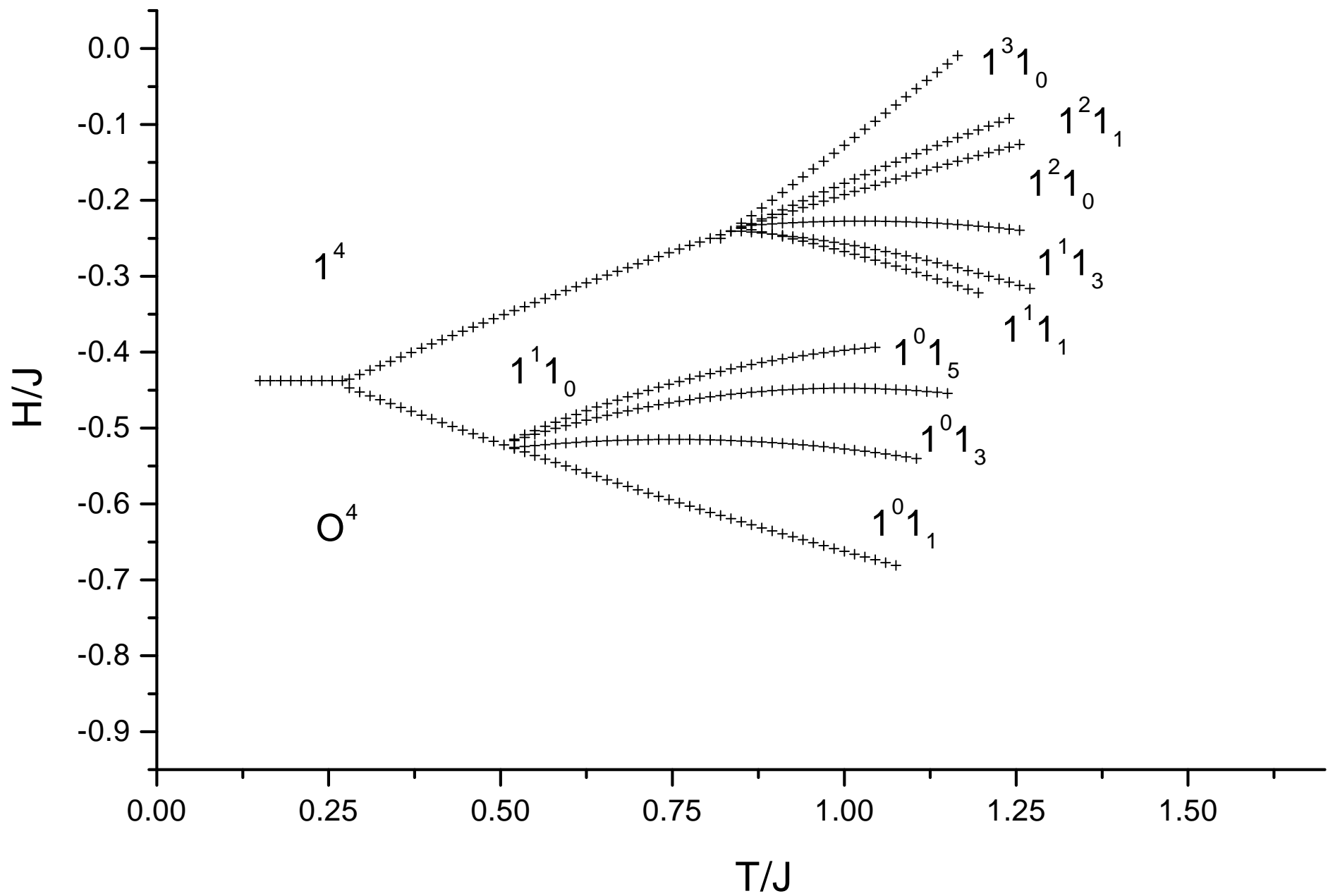
\noindent{\bf Figure 6}: Phase diagram of the intra-layering and layer-by-layer transitions, in the plane $(H/J,T/J)$, using the Monte Carlo simulations for a system with $N=4,n_y=100$ and $H_{s}/J=1.0$. \\               

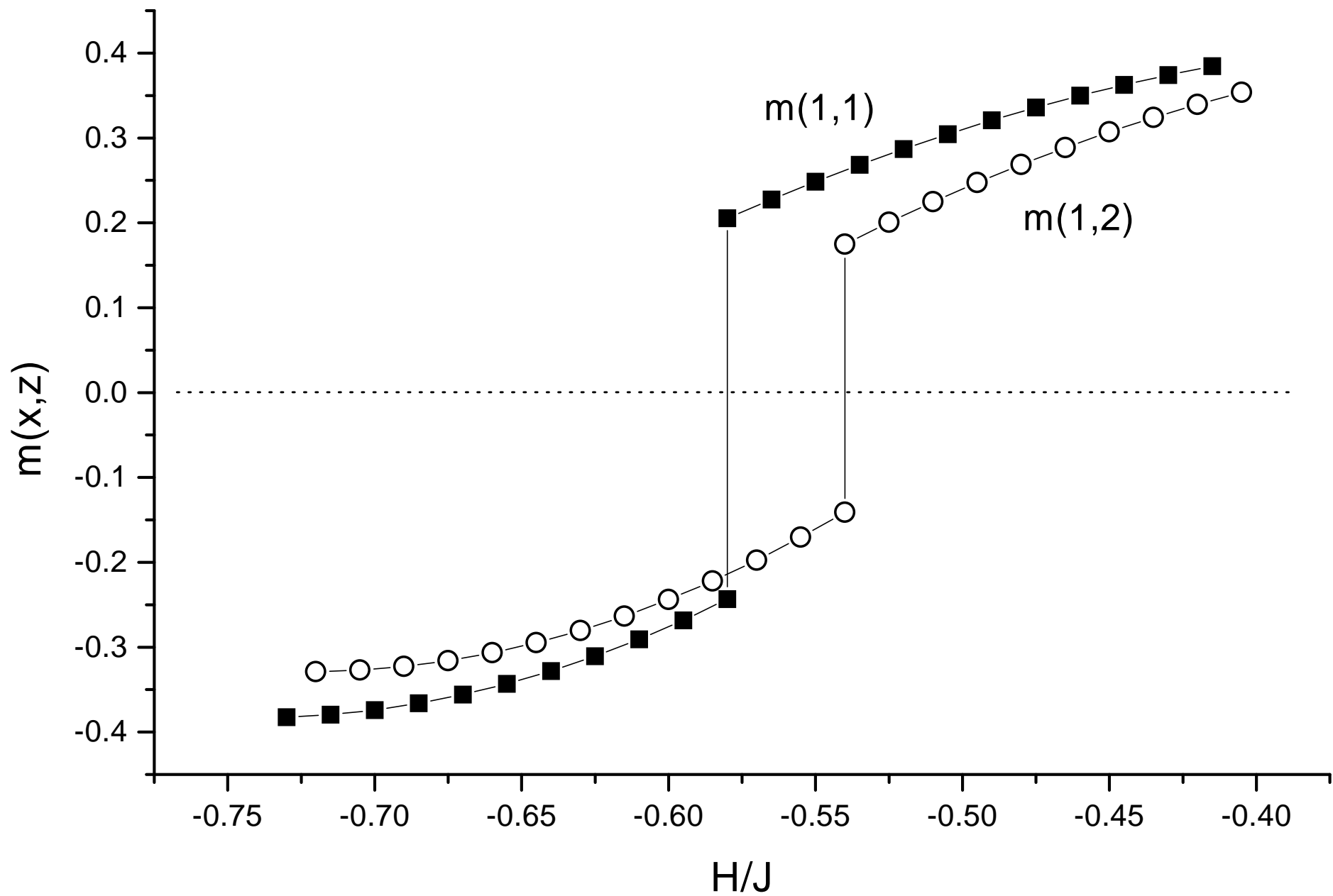
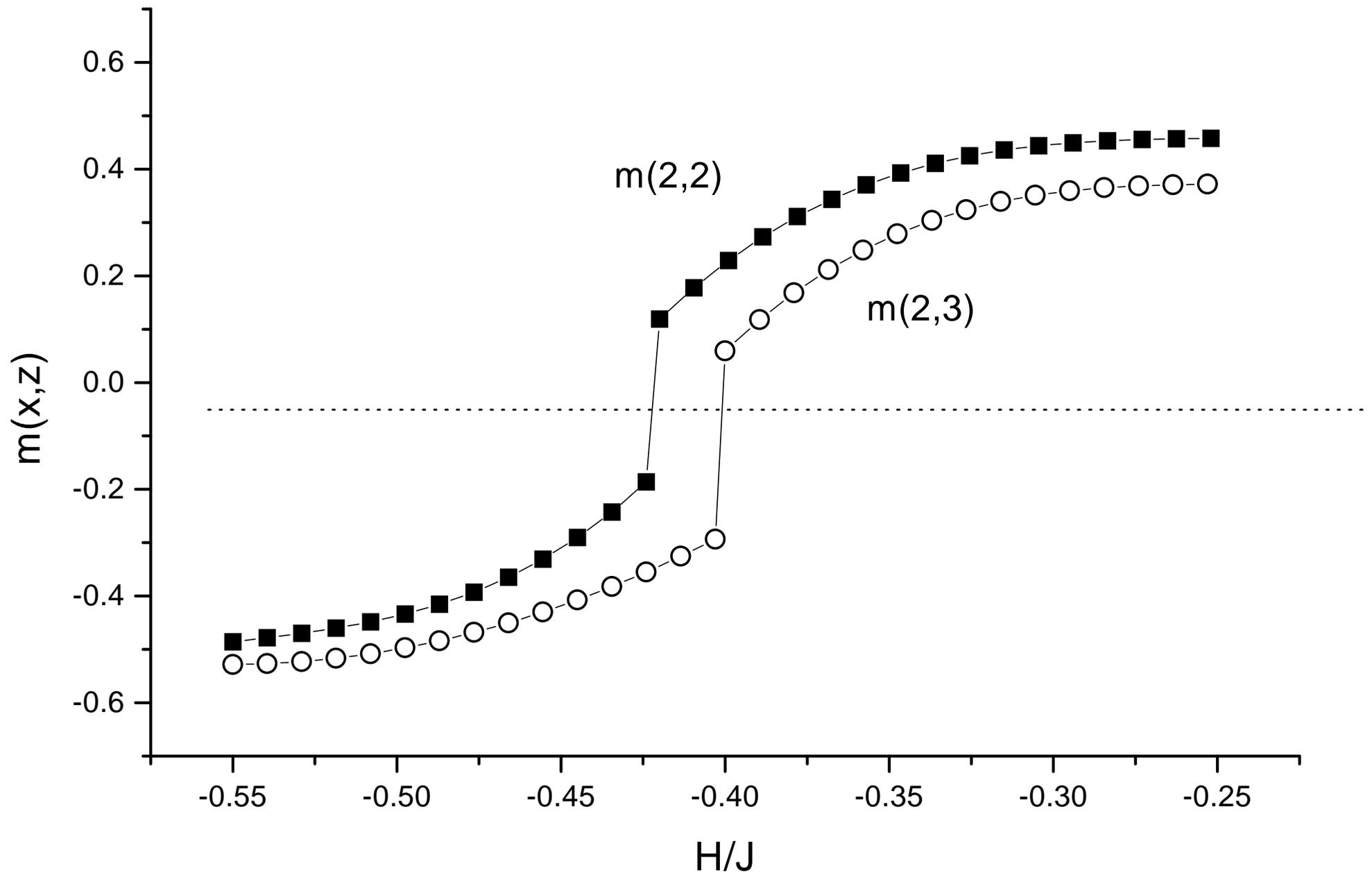
\noindent{\bf Figure 7}: Magnetization profiles as a function of the reduced bulk magnetic field $H/J$ for $N=4$ and $H_{S}/J=1.0$, of $m(1,1)$ and $m(1,2)$ at $T/J=0.75$ (a); $m(2,2)$ and $m(2,3)$ at $T/J=1.0$ (b), using the Monte Carlo method. \\

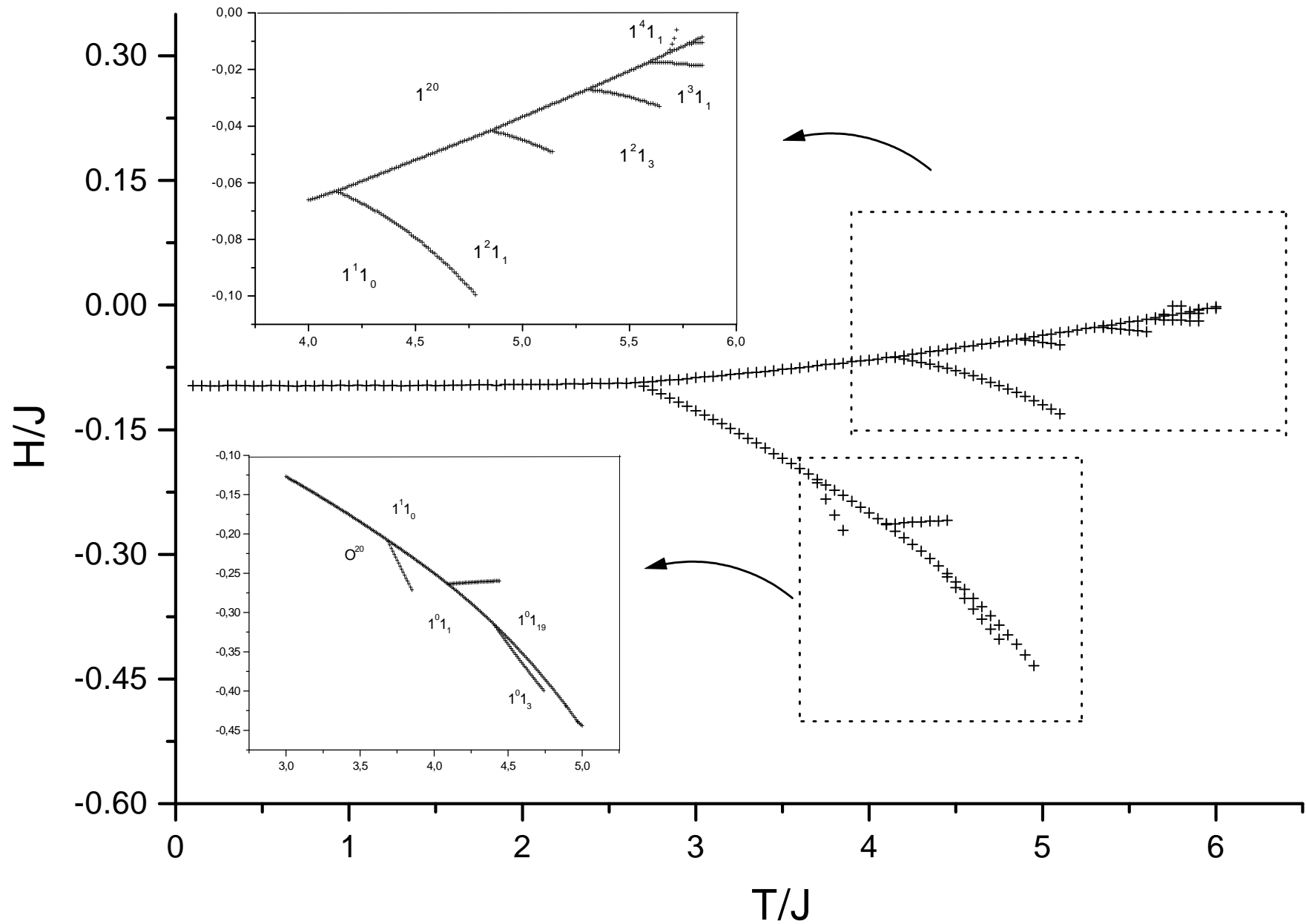
\noindent{\bf Figure 8}: Phase diagram of the intra-layering and layer-by-layer transitions, in the plane $(H/J,T/J)$, using the mean field method for a system with $N=20$ and $H_{s}/J=1.0$. \\

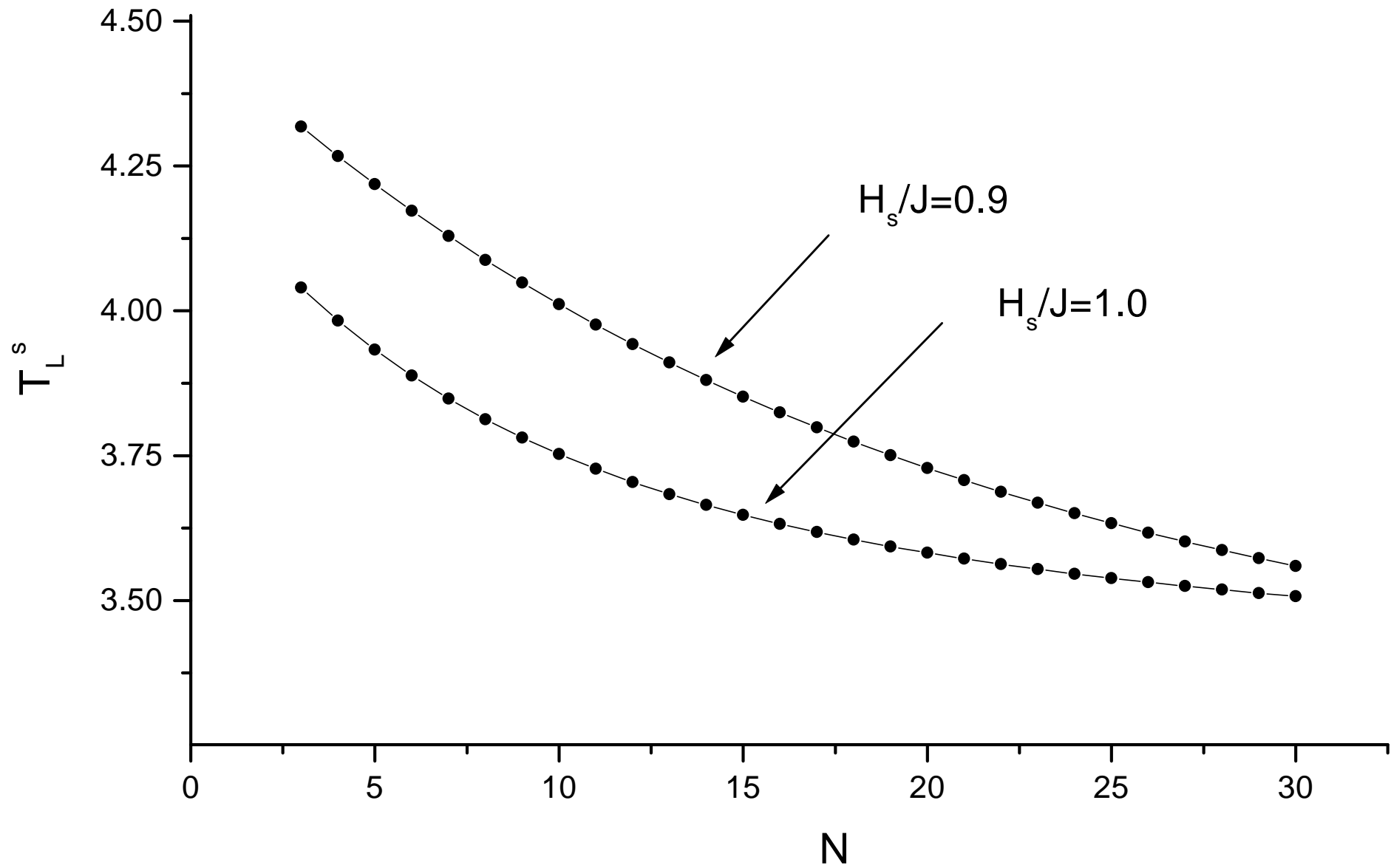
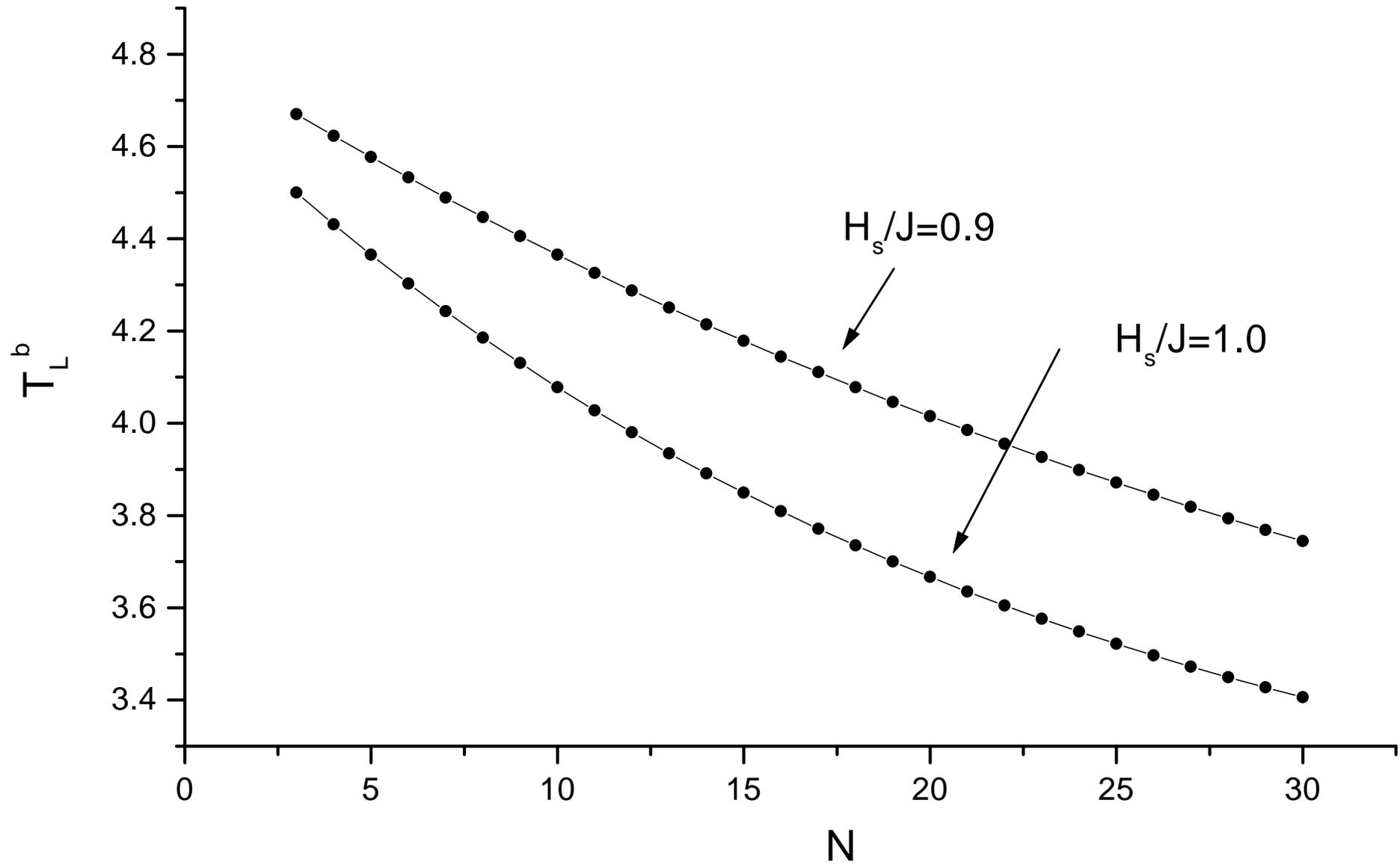
\noindent{\bf Figure 9}: Surface (a) and bulk (b) layering temperature profiles as a function of the system size $N$ for two surface magnetic field values $H_{S}/J=1.0$ and $H_{S}/J=0.9$, by using the mean field method. \\

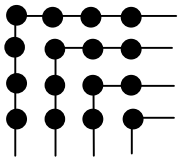
\end{document}